\documentclass[pdflatex,sn-mathphys-num]{sn-jnl}

\usepackage{array,mathtools,amssymb,booktabs,multirow,diagbox,hhline,natbib}

\usepackage{graphicx,graphics,color}
\usepackage{dcolumn}
\usepackage{bm}
\usepackage{bbold}
\usepackage{amsmath}
\usepackage{xcolor}

\allowdisplaybreaks

\newcommand{\mean}[1]{\left\langle{#1}\right\rangle}
\newcommand{\condl}{\mean{\bar q q}}

\newcommand{\gsim}{\raise.3ex\hbox{$>$\kern-.75em\lower1ex\hbox{$\sim$}}}

\begin{document}


\title{The QCD scalar susceptibility and  thermal scalar resonances in chiral symmetry restoration}

\author[1]{\fnm{A. G\'omez Nicola} }\email{gomez@ucm.es}

\author[1]{\fnm{A. Vioque-Rodr\'iguez}}\email{avioque@ucm.es}

\affil[1]{\orgdiv{Facultad de Ciencias F\'isicas, Departamento de F\'isica Te\'orica}, \orgname{Universidad Complutense de Madrid and IPARCOS}, \orgaddress{\street{Plaza de las Ciencias 1}, \postcode{28040} \city{Madrid}, \country{Spain}}}

\abstract{Building upon recent results on the role of thermal resonances in chiral symmetry restoration, we show that a description of the QCD scalar susceptibility at finite temperature $T$ saturated by the thermal properties of the lightest scalar resonance, the $f_0(500)$, is compatible both with lattice QCD data at nonzero $T$ and with the $T=0$ light resonance properties coming from experimental data. The thermal $f_0(500)$ is generated within the framework of Unitarized Chiral Perturbation Theory. This method allows us to achieve a good description of lattice QCD results with a reliable pion mass dependence. In particular, we perform direct fits to the chiral susceptibility measured in lattice data at different pion masses and temperatures, obtaining a remarkable agreement for the susceptibility and for mass differences of the light quark condensate. In addition, the fitted low-energy constants are compatible with $T=0$ phenomenology. Our results confirm the role of unitarized approaches and thermal resonances in the dynamics of the QCD transition.}
 \maketitle
\section{Introduction}
\label{secintro}
In recent years, experimental and lattice collaborations have made significant progress exploring the properties of the QCD phase diagram at finite temperature $T$ and moderately low baryon chemical potential $\mu_B$. The restoration of chiral symmetry remains one of the central aspects to understand the QCD phase diagram. Extensive studies have established that for $N_f =2+1$ flavors, with light quark masses $m_u=m_d=m_l\ll m_s$, the chiral transition is a crossover, occurring at a temperature of approximately $T_c\sim 155$ MeV \cite{Aoki:2009sc, Borsanyi:2010bp, Bazavov:2011nk,Buchoff:2013nra,Ding:2024sux}. However, a true chiral phase transition holds only in the idealized case of $N_f=2$ with massless light quarks. In this case, the results of lattice simulations suggest that the transition temperature is reduced to $T_c\sim 132$ MeV \cite{HotQCD:2019xnw}. The connection of the chiral transition with $U(1)_A$ asymptotic restoration is also a crucial challenge, related to the order of the transition and degeneration of chiral partners involving strangeness \cite{Pelissetto:2013hqa,GomezNicola:2020qxo,GomezNicola:2023rqi}.

The study of chiral symmetry restoration has focused on  two key quantities: the light quark condensate, $\langle \bar{q}q\rangle_l$, and the scalar susceptibility, $\chi_S$,  
\begin{eqnarray}
     \langle \bar{q}q\rangle_l(T)&=&\langle \bar{u}u\rangle+\langle \bar{d}d\rangle=\dfrac{\partial z}{\partial m_l},\\
     \chi_S(T)&=&-\dfrac{\partial}{\partial m_l}\langle \bar{q}q\rangle_l(T),
\end{eqnarray}
with the vacuum energy density $z(T)=-\text{lim}_{V\rightarrow \infty}(T/V)\,\text{log}\,Z$,  the QCD partition function $Z$ and $m_l=m_u=m_d$.
  
The light quark condensate is the order parameter of the transition. In the physical mass case, it develops an inflection point, while in the light chiral limit, it vanishes at the phase transition. On the other hand, the scalar susceptibility measures the order parameter correlations and diverges at $T_c$ in the light chiral limit for $N_f=2$ \cite{Smilga:1995qf}, whereas in the physical case, it develops a peak at $T_c$ \cite{Aoki:2009sc,Ding:2024sux}.

It is crucial to have theoretical approaches capable of capturing the dynamics of quark condensates and susceptibilities so that they can be later compared to lattice data. Since chiral symmetry is intrinsically connected to the low-energy sector of QCD, a consistent theoretical description of the evolution of observables below the transition must incorporate effective theories that include pions for $N_f=2$ and kaons and eta for $N_f=3$. This is particularly important for thermodynamical quantities dominated by the lightest state of the hadron gas, as it happens to be the case for $\chi_S$. In that context, Chiral Perturbation Theory (ChPT) provides a suitable framework from which  relevant observables can be built for the QCD phase diagram when combined with unitarization, thermal resonances and Ward Identities (WI) \cite{Nicola:2020smo}.

Although standard ChPT predicts results for $\langle \bar{q}q\rangle_l(T)$ and $\chi_S(T)$ consistent with chiral symmetry restoration \cite{GomezNicola:2012uc}, achieving more accurate predictions near $T_c$ usually requires including higher-mass hadronic states as $T$ increases, because of their distinctive Boltzmann weight. The most widely used approach within that framework has been the Hadron Resonance Gas model (HRG) \cite{Hagedorn:1968zz,Karsch:2003vd,Tawfik:2005qh,Huovinen:2009yb,Jankowski:2012ms}. The HRG model works quite well below $T_c$, although qualitatively it does not reproduce the expected inflection point for the quark condensate, nor the peak of the scalar susceptibility in the crossover regime, since it gives rise to a monotonically increasing function, as showed in detail in \cite{Ferreres-Sole:2018djq}.

Alternatively, it has been shown
that light thermal meson resonances, generated through unitarized ChPT, play a crucial role to describe scalar susceptibilities near $T_c$
\cite{Ferreres-Sole:2018djq,GomezNicola:2020qxo,GomezNicola:2023rqi}. Here, we will extend the analysis in \cite{Ferreres-Sole:2018djq} for the description of $\chi_S$ from the thermal $f_0(500)$, performing a detailed quantitative study based on fits of the Low-Energy Constants (LEC) of ChPT to finite-$T$ lattice data,  to show that such thermal resonance description of the chiral transition is robust and compatible with $T=0$ experimental data. The paper is organized as follows. In section \ref{secsat} we discuss the main features of the $\chi_S$ determination through saturation of the thermal $f_0(500)$ pole. In section \ref{secMchiM}, we will review the most recent observables obtained in lattice to study chiral symmetry restoration and the chiral critical temperature, and provide their ChPT expressions at finite $T$, needed in our numerical analysis.  
In section \ref{secfits}, we present our numerical results for different fits of the unitarized susceptibility to lattice values, connecting also with the light quark condensate and discussing the corresponding LEC determinations. Finally, in section \ref{secconclusions}, we will summarize our conclusions.
\section{The saturated approach}
\label{secsat}
As shown in \cite{Ferreres-Sole:2018djq}, a connection between the scalar susceptibility and the $f_0(500)$ state can be established by constructing a unitarized susceptibility,  saturated with the thermal mass of the $f_0(500)$. This relationship arises naturally because $\chi_S(T)$ is related to the correlator of $\sigma_l\sim \bar{q}_lq_l$, which has the same quantum numbers as the $f_0(500)$ and is intrinsically linked to the vacuum structure. As a testbed for such connection, one can use a Linear Sigma Model (LSM) approach \cite{Ferreres-Sole:2018djq}:
\begin{equation}
\chi_S^{LSM} (T)=4 B_0^2\left[-\dfrac{1}{2B_0 M_{0\sigma}^2}\dfrac{2M_{0\sigma}^2-3M_{0\pi}^2}{M_{0\sigma}^2-M_{0\pi}^2}\langle \bar{q}q\rangle_l(T)+\left(\dfrac{M_{0\sigma}^2}{M_{0\sigma}^2-M_{0\pi}^2}\right)^2
\frac{v^2}{M_{0\sigma}^2+\Sigma(0,\vec{0};T)}
\right],
\label{susmod}
\end{equation}
with $v$ the $T=0$ potential minimum of the LSM lagrangian, $M_{0\sigma}^2$ and $M_{0\pi}^2= 2 B_0 m_l$ the sigma and pion tree-level masses
and $\Sigma(k_0,\vec{k};T)$ the sigma self-energy \cite{Masjuan:2008cp,Manohar:2008tc,Ayala:2000px,Ferreres-Sole:2018djq}. 

As shown in \cite{Ferreres-Sole:2018djq}, the term proportional to $\langle \bar{q}q\rangle_l(T)$ in \eqref{susmod} is $\mathcal{O}(M_{0\pi}^2/M_{0\sigma}^2)$ suppressed with respect to the second term near the transition. Thus, the thermal behavior of the susceptibility is approximately described around the transition by
\begin{equation}
\dfrac{\chi_S^{LSM} (T)}{\chi_S^{LSM} (0)}\simeq \dfrac{M_{0\sigma}^2+\Sigma\left(k=0;T=0\right)}{M_{0\sigma}^2+\Sigma\left(k=0;T\right)}.
\label{susgreen}
\end{equation}
This connection between the scalar susceptibility and the self-energy around the transition in the LSM can be interpreted as a saturation approach. Despite the limitations of the LSM, the saturated susceptibility shows a chiral symmetry restoration trend and provides a reasonable description of the lattice data below $T_c$,  although it is unable to reproduce the crossover peak \cite{Ferreres-Sole:2018djq}. 

The previous ideas can be applied without the need to invoke an explicit $\sigma$ field realization, using a well-established framework for describing resonances: Unitarized Chiral Perturbation Theory (UChPT). In this context, the thermal $f_0(500)$ emerges as a pole in the second Riemann sheet  ($s_p=M_p-i\Gamma_p/2$) of the $\pi\pi$ scattering amplitude at finite $T$ \cite{Dobado:2002xf} and the susceptibility is expected to scale around $T_c$ as \cite{Ferreres-Sole:2018djq}:
\begin{equation}
    \chi_S^U(T)= A B_{phys}^2 \dfrac{M_S^2(0)}{M_S^2(T)},
    \label{susunit}
\end{equation}
where we have used the same normalization as in \cite{Ferreres-Sole:2018djq}, with $B\equiv M_\pi^2/(2m_l)$ and  "{\em phys}" meaning 
that physical values are taken for meson and quark masses and meson decay constants. In the above equation,  $M_S^2(T)=\text{Re}\,s_p(T)=M_p^2(T)-\Gamma_p^2(T)/4$,  the squared thermal mass, which  would correspond  to 
$M_{0\sigma}^2+\Sigma\left(k=0;T\right)$ in the LSM, and $A=\chi_S^U (0)/B_{phys}^2$ will be used as one of our fit parameters (see below). 
In turn, we recall that $s_p(T)$ has been also calculated within the LSM in \cite{Lyu:2024lzr}. In the LSM, the width of the sigma resonance increases at low temperatures due to the enhanced thermal phase space, and then rapidly drops to zero at a temperature below the transition. A similar qualitative trend is observed in UChPT, although the pole remains near threshold and retains a sizable width around the transition.


Using the LEC determined at $T=0$ in \cite{Hanhart:2008mx} to fit experimental phase shifts within their uncertainties, and setting $A=A_{ChPT}=\chi_S^{ChPT}(T=0)/B_{phys}^2$ in \eqref{susunit}, already gives rise to a $\chi_S$ peak around $T_c$, thus improving qualitatively the HRG in that region \cite{Ferreres-Sole:2018djq}. It is worth pointing out that the pole position parameters of the $f_0(500)$ are most sensitive to the renormalized $SU(2)$ LECs appearing in the $\pi\pi$ scattering vertices, i.e., $l_1^r$ and $l_2^r$, than to $l_3^r$ and $l_4^r$, which come from the renormalization of the pion mass and the pion decay constant.  In addition, leaving $A$ as a free fit parameter confirms this behaviour, giving a value for $A$  compatible with $A_{ChPT}$ within uncertainties. Recall that the $l_i^r$ depend on the renormalization scale, ensuring the scale independence of the observables, but they are mass-independent \cite{Gasser:1983yg}.



The same ideas can be used in the  $I=1/2$ isospin channel, where the lightest scalar state is the $K_0^*(700)$  (also known as $\kappa$). A unitarized susceptibility $\chi_S^{\kappa}$ can be constructed by saturating it with the thermal $K_0^* (700)$ pole. The results are compatible with the behaviour of reconstructed lattice $\chi_S^\kappa,\chi_P^K$ susceptibilities obtained from WI relating them to light- and strange-quark condensates,  $\chi_S^\kappa$ showing a peak above $T_c$ signaling $U(1)_A$ restoration \cite{GomezNicola:2023rqi,GomezNicola:2020qxo}. In addition, the modification of the $f_0(500)$ spectral properties and $\chi_S$ with the temperature and the axial chemical potential $\mu_5$ have also been studied within this approach  \cite{Espriu:2020dge,GomezNicola:2023ghi}. The dependence  $T_c (\mu_5)$  has been fitted to lattice values, showing an increasing behavior with $\mu_5$ consistently with lattice results. This analysis also allows to improve the determination of new LEC appearing in the $\mu_5\neq 0$ Lagrangian.
\section{Lattice observables and their ChPT expressions}
\label{secMchiM}
In the lattice, the light quark condensate at finite temperature is affected by $T = 0$ finite-size divergences. These divergences can be removed by subtracting an appropriate observable that shares the same divergences. A quantity free from finite-size divergences is the so-called subtracted quark condensate, defined as:
\begin{equation}
    \Delta_{l,s}=\langle \bar{q}q\rangle_l-2\,\dfrac{m_l}{m_s}\, \langle \bar{s}s\rangle,
\end{equation}
with $\langle \bar{s}s\rangle$ the strange quark condensate given by $\langle \bar{s}s\rangle(T)=\partial z/\partial m_s$.

\begin{figure}[h]
    \centering
        \includegraphics[width=0.42\textwidth]{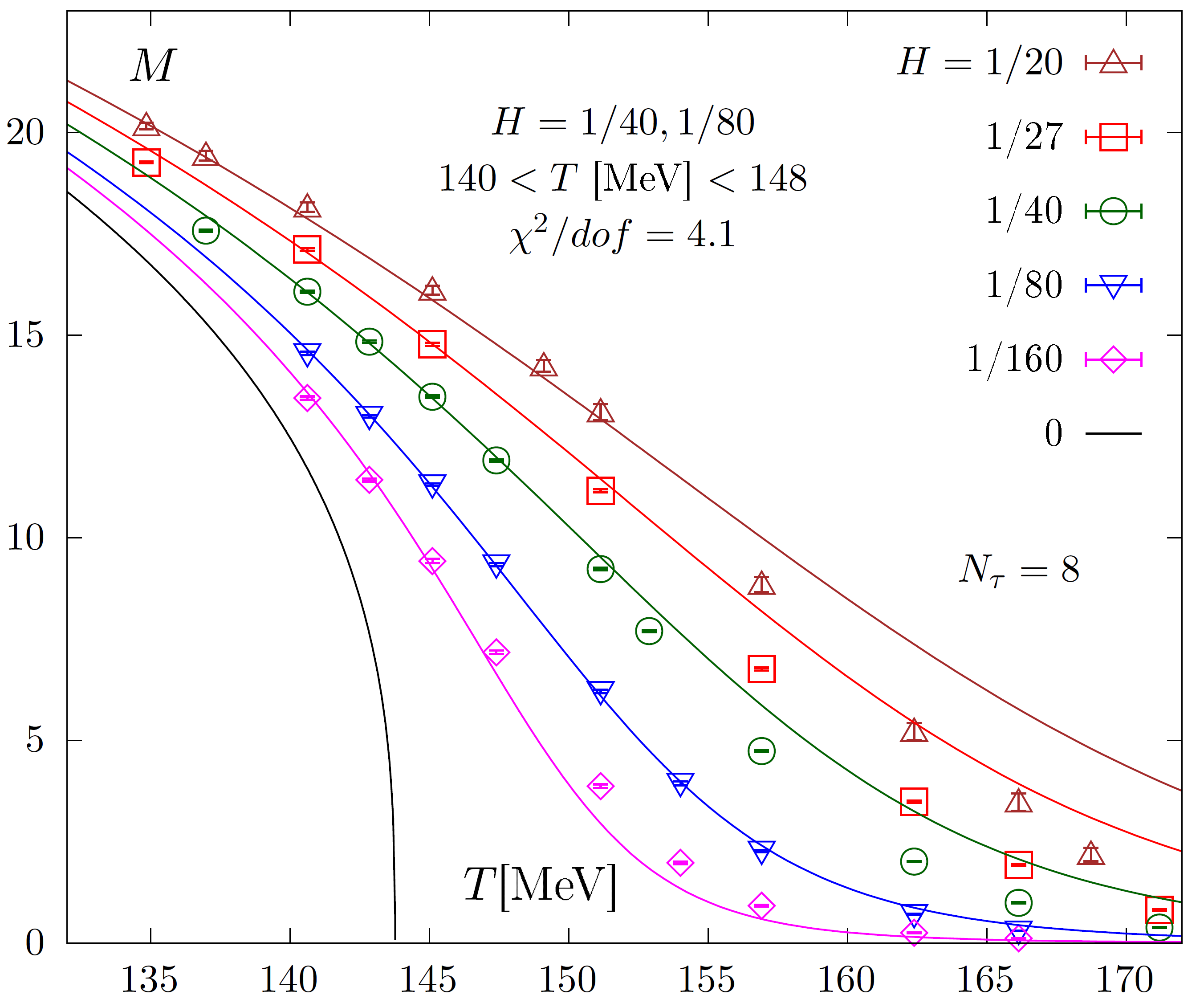}
        \includegraphics[width=0.45\textwidth]{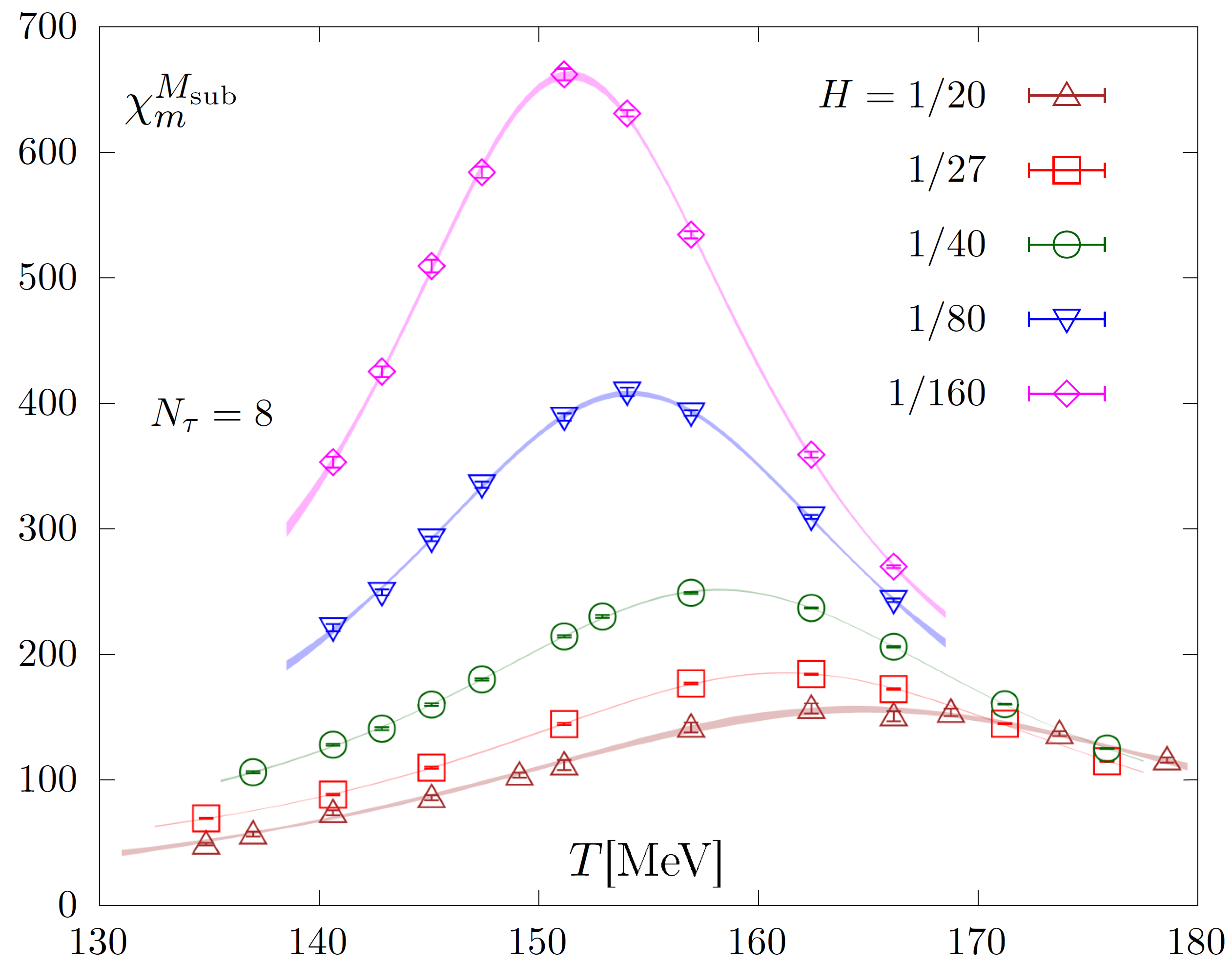}
    \caption{From \cite{Ding:2024sux}: Temperature dependence of the dimensionless subtracted quark condensate $M$ (left) and the chiral susceptibility $\chi_M$ (right) obtained on lattice for different ratios of $H=m_l/m_s$ and with temporal extent $N_\tau = 8$. }
    \label{fig:MandchiMlatt}
\end{figure}

The dimensionless order parameter $M= -m_s^{phys} \Delta_{l,s}/ (F_K^{phys})^4$ is also commonly used, where $F_K$ is the kaon decay constant. An observable of interest measured in the lattice and connected to the scalar susceptibility $\chi_S$ is the so-called chiral susceptibility $\chi_M$:
\begin{equation}
    \chi_M=m_s^\text{phys}\dfrac{\partial M}{\partial m_l}=\dfrac{(m_s^\text{phys})^2}{(F_K^\text{phys})^4}\left(\chi_S+\dfrac{2}{m_s}\langle \bar{s}s\rangle+2\dfrac{m_l}{m_s}\dfrac{\partial}{\partial m_l}\langle \bar{s}s\rangle\right).
    \label{chiMchiSmas}
\end{equation}
Around the transition temperature, the strange quark condensate $\langle \bar{s}s\rangle$ exhibits a softer temperature dependence than $\langle\bar{q}q\rangle_l$ due to the explicit chiral symmetry breaking $m_s\gg m_l$. Consequently, the quantity $M$ keeps the main features of the light quark condensate and $\chi_M$ those of the scalar susceptibility. In Fig. \ref{fig:MandchiMlatt} we show the lattice results for those quantities calculated in \cite{Ding:2024sux}. Around $T_c$, the inflection point of $M$ and the peak of $\chi_M$ are clearly observed, and they move to lower temperatures as the parameter $H=m_l/m_s$ is reduced, which is their expected trend.


Therefore, we see that the relevant $T$-dependence of $\chi_M$ around $T_c$ is dominated by that of $\chi_S$, so that the remaining terms on the right-hand side of \eqref{chiMchiSmas} are expected to be relatively small around the transition. Our next step will be to justify that assumption from the ChPT expressions of those quantities, which in turn will help us to construct our relevant fitting functions, including the unitarized $\chi_S(T)$ in \eqref{susunit}.
For that purpose, we will make use of the $SU(3)$ ChPT expressions of $\chi_S$ and $\langle \bar{s}s\rangle$  calculated to NLO in the chiral power counting at $T\neq 0$ in \cite{GomezNicola:2012uc} in terms of the renormalized  mass-independent $SU(3)$ LEC ($L_i^r$,$H_i^r$) and $M_{\pi,K,\eta}$.  We get 
\begin{eqnarray}
\Delta\chi_S(T)&=&2B_0^2\left[3g_2(M_\pi,T)+g_2(M_K,T)+\dfrac{1}{9}g_2(M_\eta,T)\right]+\mathcal{O}\left(\dfrac{1}{F^2}\right),\\ \nonumber
    \tilde\Delta\chi_M(T)&=&2B_0^2\left\lbrace 3g_2(M_\pi,T)+\left( 1-2\dfrac{m_l}{m_s}\right)g_2(M_K,T)+\frac{1}{9}\left(1-4\dfrac{m_l}{m_s}\right)g_2(M_\eta,T)\right.\\ 
    &-&\left.\dfrac{2}{m_sB_0}\left[g_1(M_K,T)+\frac{1}{3}g_1(M_\eta,T)\right]\right\rbrace+\mathcal{O}\left(\dfrac{1}{F^2}\right),
\end{eqnarray}
where we have parametrized the chiral expansion in  powers $F^{-n}$ with $F$ the pion decay constant in the chiral limit, the $g_{i}$ functions are given in \cite{GomezNicola:2012uc}, $\Delta\chi_S(T)=\chi_S(T)-\chi_S(0)$ and $\tilde\Delta\chi_M(T)=\dfrac{(F_K^\text{phys})^4}{(m_s^\text{phys})^2}\left[\chi_M(T)-\chi_M(0)\right]$.

The kaon and eta thermal functions $g_{i}(M_{K,\eta}, T)$ are Boltzmann suppressed with respect to the pion thermal loops in the $T \ll M_K$ regime. In addition, $g_2(M_\pi,T)$ behaves proportionally to $T/M_\pi$ near the chiral limit $M_\pi\rightarrow 0^+$. Thus, the pion contribution dominates both $\chi_S$ and $\chi_M$, and $\tilde\Delta\chi_M(T)\simeq \Delta\chi_S(T)$ is a good approximation in ChPT around the transition. In the following, we will use that approximation,  supported by both ChPT and the lattice results. Thus, we will assume that the possible differences between $\tilde\Delta\chi_M(T)$ and $\Delta\chi_S(T)$ lie within the typical uncertainty range of the saturated approach. From our previous comments, we expect this approach to provide the expected peak behaviour around $T_c$ with $SU(2)$ LEC compatible with $T=0$ and in fact, our fits in the next section will confirm quantitatively this claim. 
\section{Fits and numerical results}
\label{secfits}
The advantage of the UChPT approach is that is built from the ChPT amplitude, which by construction bears the correct quark and pion mass dependence up to a given order in the chiral expansion. In particular, we have access to the pole parameters of the $f_0(500)$, and hence of $\chi_S(T)$, as a function of $M_\pi$, which is particularly useful for the comparison with lattice data for different masses, as we are about to see. 

Based on our discussion in section \ref{secMchiM}, we will follow the following fit strategy to describe lattice results for $\chi_M$ around $T_c$. First, the thermal behaviour around $T_c$ will be controlled by the saturated approach \eqref{susunit}, bearing in mind that the normalization factor $A$ depends on the pion mass. In that context, recall that although fixing $A=A_{ChPT}$ ensures the $T=0$ matching within ChPT,  the saturated approach does not necessarily describes the behaviour for temperatures well below $T_c$, where additional contributions, such as those in the LSM expression \eqref{susmod}, could be important. For that reason, we will also treat $A$ as a fit parameter, in addition to the LEC.  Second, we will rely on the approximation  $\tilde\Delta\chi_M(T)\simeq \Delta\chi_S(T)$ around $T_c$, fixing the $T=0$ quantities with ChPT to NLO. Thus, we construct our unitarized chiral susceptibility $\chi_M^U$ as 
\begin{equation}
    \dfrac{(F_K^\text{phys})^4 (H^{\text{phys}})^2}{(M_\pi^{\text{phys}})^4}\chi_M^U(T,M_\pi)=\dfrac{A(M_\pi)}{4}\dfrac{M_S^2(0,M_\pi)}{M_S^2(T,M_\pi)}+\dfrac{1}{4B_{phys}^2}\tilde \Delta\chi_{MS}^{ChPT},
    \label{DeltachiSU}\\
\end{equation}
with
{\small
\begin{eqnarray}
\tilde\Delta\chi_{MS}^{ChPT}&=&\dfrac{(F_K^\text{phys})^4}{(m_s^\text{phys})^2}\chi_M^{ChPT}(0)-\chi_S^{ChPT}(0)=-\dfrac{2 B_0}{m_s}\left(F^2-\frac{11}{144 \pi^2} M_{0 \pi}^2\right)\nonumber \\
&+&\dfrac{2 B_0}{m_s}\left\lbrace \frac{1}{144 \pi^2} \left[ 
   9 \left( 2 M_{0 K}^2 + M_{0 \pi}^2 \right) \log \left( \frac{M_{0 K}^2}{\mu^2} \right) + 
   2 \left( 3 M_{0 \eta}^2 + M_{0 \pi}^2 \right) \log \left( \frac{M_{0 \eta}^2}{\mu^2} \right) \right]\right.\nonumber\\
   &+& \left. 4\left[-2 (H_2^{r} + 4 L_6^{r} + 2 L_8^{r}) M_{0 K}^2 + (H_2^{r} + 2 (-6 L_6^{r} + L_8^{r})) M_{0 \pi}^2\right]\right\rbrace+\mathcal{O}\left(\dfrac{1}{F^2}\right).
\end{eqnarray}}

\begin{figure}[h!]
    \centering
    \includegraphics[width=0.5\textwidth]{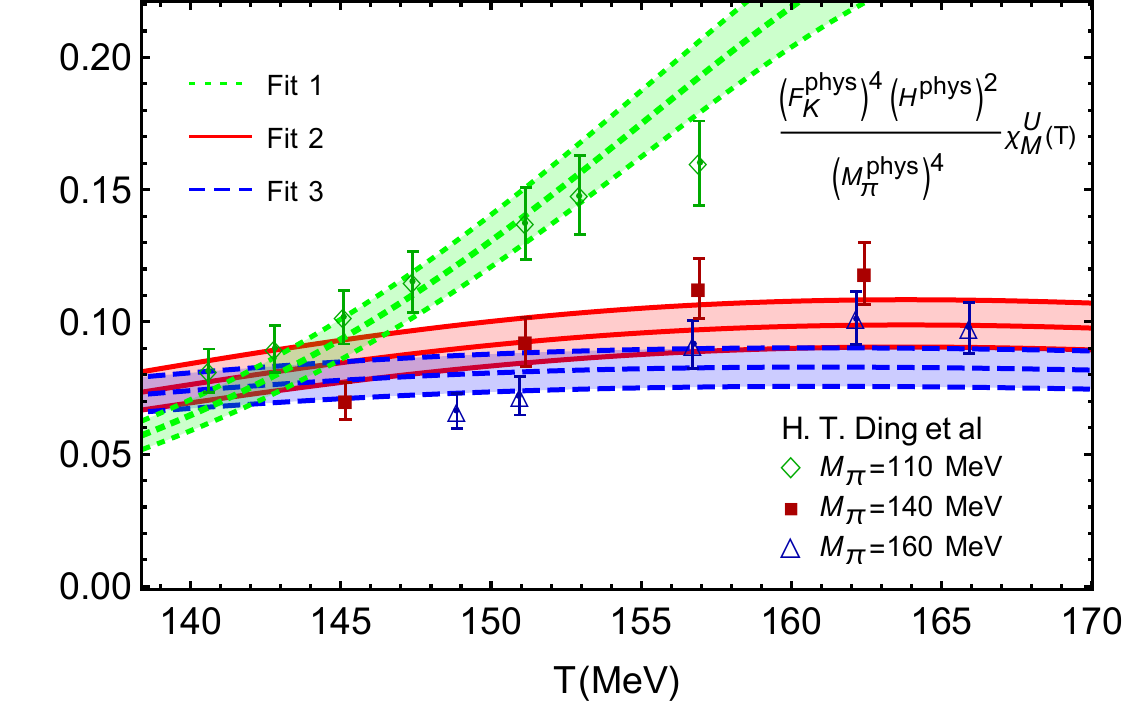}
    \includegraphics[width=0.49\textwidth]{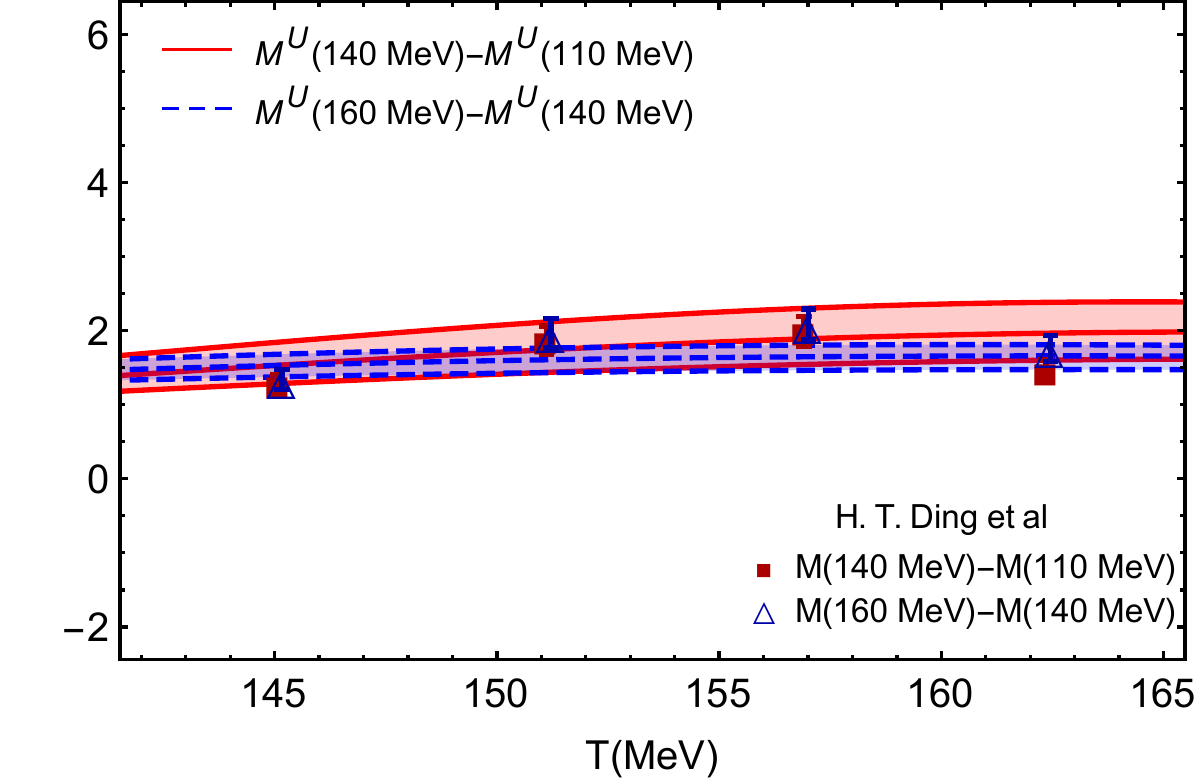}
    \caption{Left: fit 1, 2, and 3 of the $f_0(500)$ saturated scalar susceptibility. Right: difference between the dimensionless order parameter $M$ evaluated at two different pion masses using the constants obtained from fit 4. A characteristic uncertainty of $10\%$ has been assigned to the lattice data corresponding to the $M$ differences. The uncertainty bands correspond to the uncertainties in the fit parameters.}
    \label{fig:susMfit}
\end{figure}

As mentioned above, one of our purposes here is to fit the $l_{i}^r$ with finite-$T$ lattice results and is reasonable to use only $l^r_{1,2}$ as fit parameters, whereas we fix  $l^r_{3,4}$ to their original ChPT values \cite{Gasser:1983yg}, which is the same procedure followed in \cite{Hanhart:2008mx} where $l_{1,2}^r$ were obtained by a fit of the IAM to scattering data. We show here the results of several fits,  summarized in Table \ref{fittab} and Figure \ref{fig:susMfit}. Consequently, with our $N_f=2$ approach at finite $T$, we will fix $m_s$ and translate lattice data at different values of $H=m_l/m_s$ (see Fig.\ref{fig:MandchiMlatt}) to pion masses $M_\pi=160, 140$ and $110$ MeV. For the numerical computation, we are considering $M_\pi^{\text{phys}}=140\,\text{MeV}$, $F_\pi^{\text{phys}}=93\,\text{MeV}$, $F_K^\text{phys}=110.4\,\text{MeV}$ and $m_l^{\text{phys}}=5.5\,\text{MeV}$.
For the numerical values of the $L_i^r$ we take the results from \cite{Molina:2020qpw}. In addition, we have assigned a characteristic uncertainty of $10\%$ to the lattice data. On the other hand, as the saturated approach is especially adequate near the transition region, we have included only the lattice points located around the peaks of the lattice data, which are plotted in Figure \ref{fig:susMfit}. For each pion mass (see below), these points are located within a temperature range between the value closest to the critical temperature reported in \cite{Ding:2024sux} and 
a few MeV below that value. 

As a first analysis, we fix the $l_{1,2}^r$  constants to the values given in \cite{Hanhart:2008mx} and examine the $\chi_M^U$   peak position, which is independent of $A$. For this set of LEC, the susceptibility near the physical pion mass reproduces the expected peak at $T_c$, as shown already in \cite{Ferreres-Sole:2018djq}. In fact, we obtain critical temperatures consistent with the expected values: $T_{c}^{140}=163.8\pm 5.8\, \text{MeV}$ and $T_{c}^{160}=160.9\pm 5.7\, \text{MeV}$ where the uncertainties come from those in $l^r_{1,2}$. However, for $M_\pi = 110\,\text{MeV}$, the susceptibility becomes nearly divergent, and the peak appears at a significantly higher temperature, in contrast to what is observed in lattice data. This is a consequence of our $N_f=2$ method to reach the chiral limit, which does not exactly replicate the procedure used in lattice simulations where the ratio $m_l/m_s$ is varied. Nevertheless, although our approximation does not reproduce the peak at low values of $M_\pi$, it provides a reasonably good description below and close to the peak. This sets naturally the lower pion mass limit for the applicability range of our current analysis. With the above comments in mind, we have performed  fits 1, 2 and 3 of 
the unitarized chiral susceptibility 
\eqref{DeltachiSU} with $l_{1,2}^r$ fixed and using only the normalization constants $A(M_\pi)$ as  fit parameters. Note that the value of $A(M_\pi^\text{phys})$ quoted in Table \ref{fittab} is compatible with that obtained in \cite{Ferreres-Sole:2018djq}, which was derived using different lattice results \cite{Aoki:2009sc}. 
The results shown in Figure \ref{fig:susMfit} agree with lattice data near the transition, including the expected growth of $\chi_M^U$ when the pion mass is reduced. 

In order to incorporate  $l_{1,2}^r$ as fit parameters, we will first explore the dependence on these parameters of $T_c$ (independent of the $A$ parameter) where $\chi_M^U$ reaches its maximum.  The first important observation is that the $IJ=00$ partial wave depends only on the combination $l_2^r+1.5 \,l_1^r$ in the chiral limit and therefore we expect a correlation between $l_1^r$ and $l_2^r$ within that range for $T_c$, corrected for the physical pion mass. Actually, imposing the physical constraint $157\,\text{MeV} < T_{\text{max}} < 166\,\text{MeV}$ for $M_\pi=140\, \text{MeV}$, so that $T_c$ remains  around its physical value for $l_{1,2}^r$ fixed within uncertainties, we find the approximate correlation $l_2^r+1.6 \,l_1^r\equiv l_{2,0}^r$
with $-1.6\lesssim l_{2,0}^r\times 10^3\lesssim -0.7$. 
With those constraints, we carry out a combined fit (fit 4) using the same lattice data as in the previous fits for the three different pion masses, keeping the mean values of $A$ obtained in those fits fixed, using as fit parameters  $l_1^r$ and $l_{2,0}^r$. 
The resulting $l_{1,2}^r$ are compatible with those in \cite{Hanhart:2008mx} and other determinations \cite{Doring:2016bdr}, although with large uncertainties. The uncertainty is reduced considerably when $l_{2,0}^r$ is the only fit parameter (fit 5), confirming that this parameter provides the main LEC dependence for the susceptibility. 
In addition, with fit 4 values, we obtain a reasonable pole position for the $f_0(500)$ resonance at $T=0$, namely  $\sqrt{s_p}^{f_0}=(449^{+4}_{-8})-i\,(222^{+9}_{-13}) \,\text{MeV}$ compatible with the PDG \cite{ParticleDataGroup:2024cfk}. However, that is not the case for the $\rho(770)$ resonance. 
This can be improved by considering a combined fit (fit 6) with the five free parameters $l_1^r,l_{2,0}^r, A(110),A(140),A(160)$ and demanding that the $\rho(770)$ pole mass and width are also fitted within a characteristic uncertainty of $10\%$ around their PDG values $\sqrt{s_p\,}^{\rho}=(775.3- 73.9\, i)$ MeV \cite{ParticleDataGroup:2024cfk}.  
Such uncertainty range still allows for a good description of the finite-$T$ lattice results. As a result, we obtain a more reliable determination of the LEC uncertainties as well as a better $\chi^2$/dof for this fit (see Table \ref{fittab}). 
For this fit, $T=0$  pole position for the $f_0(500)$ is  $\sqrt{s_p}^{f_0} = (447.8^{+0.3}_{-1.5}) - i(217.0^{+3.9}_{-0.6}) \,\text{MeV}$.

Finally, we discuss how to extract information about the subtracted quark condensate from our unitarized susceptibility approach. For that purpose, we rely on the pion mass dependence of  $\chi_M^{U}(T,M_\pi)$, interpolating smoothly 
$A(M_\pi)$  for $110 \,\text{MeV}\lesssim M_\pi \lesssim 160\,\text{MeV}$ from the $A$ values obtained in the fits for the three pion masses. Then, from equation \eqref{chiMchiSmas}, we get
\begin{equation}
    M^U(T,M_\pi(m_l))-M^U(T,M_\pi(m_l^0))=\dfrac{1}{m_s}\int_{m_l^{0}}^{m_l}\,dm'_l \,\chi_M^U[T,M_\pi(m'_l)],
\end{equation}
fot $\,m_l>m_l^0$. 
This quantity is plotted in Figure \ref{fig:susMfit} for the fit 4 parameters, which 
simplify the calculation of the uncertainty bands, encoded only in the $l_i^r$ for that fit. Similar results are expected for other fits. Our results are in agreement with lattice data, which provides further evidence that the pion mass and temperature dependences of the saturated susceptibility are consistent.
\begin{table}[h!]
\caption{Results obtained from the fits discussed in the main text. The values of the LEC are given at the $\mu = 770\, \text{MeV}$ scale. The uncertainties correspond to the $95\%$ confidence level of the fit. Parameters that were not fitted but instead predicted from $l_1^r$ and $l_{2,0}^r$ are highlighted to distinguish them from the free parameters used in the $\chi^2$ minimization.} 
\begin{tabular}{c c c c c c c c c }
\toprule
 & $l_1^r\times10^3$ & $l_{2,0}^r\times10^3$ & $l_2^r\times10^3$ & $A(160)$ & $A(140)$ & $A(110)$ & $\chi^2/\text{dof}$ \\ 
\midrule
fit 1 & $-3.7$ & -0.9 & $5.0$ & - & - & $0.061^{+0.004}_{-0.003}$ & 1.6 \\
fit 2 & $-3.7$ & -0.9 & $5.0$ & - & $0.13\pm 0.01$ & - & 3.0 \\  
fit 3 & $-3.7$ & -0.9 & $5.0$ & $0.17\pm 0.01$ & - & - & 3.1 \\ 
fit 4 & $-2^{+3}_{-6}$ & $-0.8^{+0.1}_{-0.2}$&  $\bm{2^{+9}_{-5}}$ & $0.17$ & $0.13$ & $0.061$ & 2.2 \\ 
fit 5 & $-3.7$ & $-0.91\pm 0.05$&  $\bm{5.01\pm 0.05}$ & $0.17$ & $0.13$ & $0.061$ & 2.1 \\ 
fit 6 & $-3.5\pm 0.3$ & $-0.7^{+0.0}_{-0.2}$&  $\bm{4.9\pm 0.5}$ & $0.17\pm 0.01$ & $0.14\pm0.01$ & $0.073^{+0.008}_{-0.011}$ & 1.5 \\ 
\cite{Hanhart:2008mx} & $-3.7\pm 0.2$ & $-0.9\pm 0.5$ & $5.0\pm 0.4$ &  &  &  &  \\ 
\botrule
\end{tabular}
\label{fittab}
\end{table}
\section{Conclusions}
\label{secconclusions}
In this work, we have provided a quantitative analysis showing that an approach for the scalar susceptibility based on the saturation of the thermal $f_0(500)$ is capable to describe both finite $T$ lattice results around the QCD transition and $T=0$ meson phenomenological data.  
 For that purpose, we have combined unitarized ChPT describing thermal resonances and perturbative ChPT, in order to connect properly with the thermal dependence of lattice observables near chiral symmetry restoration. 

Our ChPT-based approach provides a reliable pion mass dependence, which allows us to perform fits to lattice data sets for different values of $m_l/m_s$, for which the LEC can be used as fit parameters.  The LEC obtained are consistent with the values reported in the $T=0$ literature and yield a good determination of the $f_0(500)$ and $\rho(770)$ poles, while being compatible with lattice data around $T_c$. Finally, within the same approach, we have shown that mass differences of the subtracted light condensate can be consistently described,  supporting further the pion mass and temperature dependences emerging from the saturated thermal resonance approach at the crossover region.
We remark that an accurate analytical description of both the light condensate and the scalar susceptibility is also very interesting from the viewpoint of the degeneration of the pion-like $\bar q \gamma_5 \tau^a q$ and $\sigma$-like $\bar q q$ quark bilinears, since the $\pi$ susceptibility and  the quark condensate are connected through the WI $\chi_\pi=-\condl/m_l$ \cite{Nicola:2020smo}.

\section*{Acknowledgments}

Work supported by the Ministerio de Ciencia e Innovación, research contract PID2022-136510NB-C31 and the EU Horizon 2020 program under grant agreement No 824093.

\end{document}